\documentclass[twoside]{article}
\usepackage{graphicx}
\usepackage{fancyhdr}
\usepackage{multicol}
\usepackage{styl-ap}

\begin{document}
\title{IceCube Observatory: Neutrinos and the Origin of Cosmic Rays}
\author{Paolo Desiati\work{1,2}, for the IceCube Collaboration\work{3}}
\workplace{Wisconsin IceCube Particle Astrophysics Center (WIPAC), University of Wisconsin, Madison, WI 53706, U.S.A.
\next
Department of Astronomy, University of Wisconsin, Madison, WI 53706, U.S.A.
\next
{\tt http://icecube.wisc.edu}}
\mainauthor{desiati@icecube.wisc.edu}
\maketitle

\begin{abstract}%
The completed IceCube Observatory, the first $km^3$ neutrino telescope, is already providing the most stringent limits on the flux of high energy cosmic neutrinos from point-like and diffuse galactic and extra-galactic sources. The non-detection of extra-terrestrial neutrinos has important consequences on the origin of the cosmic rays. Here the current status of astrophysical neutrino searches, and of the observation of a persistent cosmic ray anisotropy above 100 TeV, are reviewed.
\end{abstract}

\keywords{Neutrinos - Cosmic Rays - Anisotropy}

\begin{multicols}{2}

\section{Introduction}
\label{sec:intro}

One hundred years after their discovery, the origin of the cosmic rays is still a mystery. The current leading model is that cosmic rays are accelerated in diffusive shocks. In this case Supernova Remnants (SNRs) in our Galaxy could be the major source of cosmic rays up to about $10^{15}-10^{17}$ eV. The SNR energy output in the Galaxy can provide the energy budget necessary to maintain the presently observed population of galactic cosmic-rays. In particular, in order to achieve such high energies it is expected that acceleration occurs during the relatively short period in the SNR evolution between the end of free expansion and the beginning of the so-called Sedov phase. This period is about 10$^3$ years from the explosion when the shock velocity is high enough to allow for efficient acceleration. At energies in excess of about $10^{17}$ eV, Active Galactic Nuclei (AGN) and Gamma Ray Bursts (GRB) could play an important role in the origin of the extra-galactic cosmic rays. 

Since cosmic rays are deflected by magnetic fields, it is not possible to associate them to their sources. However, if hadronic particles are accelerated, a fraction of them would interact within their sources or in surrounding molecular clouds to produce mesons. The mesons eventually decay into high energy $\gamma$ rays and neutrinos with an energy spectrum $\sim E^{-2}$ of the accelerated cosmic rays. The remaining hadronic particles propagate until their detection on Earth. Detection of $\gamma$ rays and neutrinos from individual galactic or extra-galactic source candidates of cosmic rays, or from extended molecular clouds, is therefore a method to indirectly probe the origin of cosmic rays.

During the last decade, detection of $\gamma$ rays from galactic sources has been successfully achieved by satellite experiments such as AGILE and Fermi up to 10 and 100 GeV, respectively. Imaging Cherenkov Telescope Arrays such as MAGIC, VERITAS and H.E.S.S., and water Cherenkov detectors such as Milagro have made measurements up to O(10 TeV). High energy direct emission from old SNRs appears to be inconsistent with hadronic acceleration\footnote{most probably SNR older than several thousand years no longer efficiently accelerate cosmic rays.}. It is interesting, however, that delayed secondary $\gamma$ ray emissions can be produced by the most energetic particles that escaped the acceleration region when they propagate through molecular clouds that surround the star forming regions~\cite{gabici}. With this mechanism, indirect evidence of hadronic acceleration is present even when SNR are several 10$^4$ years old. In fact, the detection of an extended emission of TeV $\gamma$ rays from the Galactic Center by H.E.S.S., which is attributed to cosmic rays accelerated by SNR G0.9+0.1 interacting with the surrounding clouds, might provide the first evidence of hadronic acceleration~\cite{aharonian}. The most compelling evidence currently comes from low energy $\gamma$ ray emission from the regions surrounding the intermediate-age SNR W44. AGILE observations in the energy range of 50 MeV - 10 GeV~\cite{agile} and Fermi observations up to 100 GeV~\cite{fermi} show that while leptonic models fail to describe simultaneously $\gamma$ and radio emissions without requiring too large circumstellar densities, the hadronic models are consistent with experimental constraints from radio, optical, X and $\gamma$ rays observations. Although the $\gamma$ ray energy spectrum is consistent with a proton spectral index of 3 and a low energy cut-off of approximately 10 GeV~\footnote{this is the reason why such a source was not observed at TeV energy.}, the hadronic origin of the observed emission is considered likely. The observed steep spectrum and low energy cut-off may be caused by suppression of efficient particle acceleration in the dense environment of this source~\cite{uchiyama}. Ion-neutral collisions in the weakly ionized dense gas surrounding the remnant lead to a softer spectrum as well as to damping of the plasma Alfv\'en waves that form the shock. The resulting poor particle confinement leads to a low energy cutoff~\cite{malkov}.

Other than the specific properties of single objects, evidence of an instance of hadronic acceleration is a very important step towards the discovery of the origin of cosmic rays. However, this would not mean that all galactic cosmic rays are necessarily accelerated in SNR. If cosmic ray acceleration occurs predominantly on a larger scale, such as in superbubbles~\cite{butt} or in the Galaxy cluster medium where particles could be accelerated to ultra-high energies~\cite{lazabrunetti}, the search for the origin of cosmic rays should concentrate on extended sources or diffuse fluxes.

While the TeV $\gamma$ ray horizon is limited within our Galaxy, because of absorption in the infrared and microwave cosmic background, the GeV $\gamma$ emissions can be observed within about 100 Mpc making it possible to search for extragalactic sources of cosmic rays. On the other hand, detection of neutrinos from individual sources are an efficient and unambiguous probe for the high energy hadronic acceleration mechanism, and therefore for the sources of cosmic rays. However, the very same property that makes neutrinos an excellent cosmic messenger also makes them difficult to detect. Thus large instrumented volume of target matter is required to capture sufficient event statistics.

The IceCube Neutrino Observatory (see Fig.~\ref{fig:icecube}), completed in December 2010, is currently the only km$^3$ scale neutrino telescope collecting data. The observatory consists of an array of 5,160 optical sensors arranged along 86 cables (or strings) between 1,450 and 2,450 meters below the geographic South Pole, where the antarctic ice is particularly transparent.  IceCube includes a surface shower array, IceTop, and a dense instrumented core with a lower energy threshold, DeepCore. The surface array, IceTop, is 81 stations each consisting of two tanks of frozen clean water with each tank containing two optical sensors. IceTop, using events in coincidence with the deep IceCube array, provides the measurement of the spectrum and mass composition of cosmic rays at the knee and up to about 10$^{18}$ eV. The DeepCore sub-array, consisting of 6 densely instrumented strings located at the bottom-center of IceCube, lowers the observatory neutrino energy threshold to about 10 GeV. DeepCore uses the surrounding IceCube instrumented volume as a veto for the background of cosmic ray induced through-going muon bundles, thus enhancing the detection of down-going neutrinos within the Deep Core volume. Veto rejection power in excess of 10$^8$ has been achieved~\cite{veto}. The basic detection component of IceCube is the Digital Optical Module (DOM) which consists of a 10-inch Hamamatsu photomultiplier tube (PMT) and its own data acquisition (DAQ) circuitry enclosed in a pressure-resistant glass sphere. The DOMs detect, digitize and timestamp the signals from Cherenkov radiation photons. Their main DAQ board is connected to the central DAQ in the IceCube Laboratory at the surface, where the global trigger is determined~\cite{dom}. The construction of IceCube started in 2004 and physics quality data taking commenced in 2006. With this early data the observatory is providing the most stringent limits on the flux of high energy neutrinos from extra-terrestrial origin, and therefore strong constraints on the models of individual sources of cosmic rays and unidentified diffuse sources. At the same time, IceCube has accumulated a large number of cosmic ray induced neutrinos produced in the atmosphere, making it possible to probe the combined effect of hadronic interaction models, cosmic ray spectrum and composition on the neutrino spectrum up to a few hundred TeV~\cite{fedynitch}. 
\begin{myfigure}
\centerline{\resizebox{80mm}{!}{\includegraphics{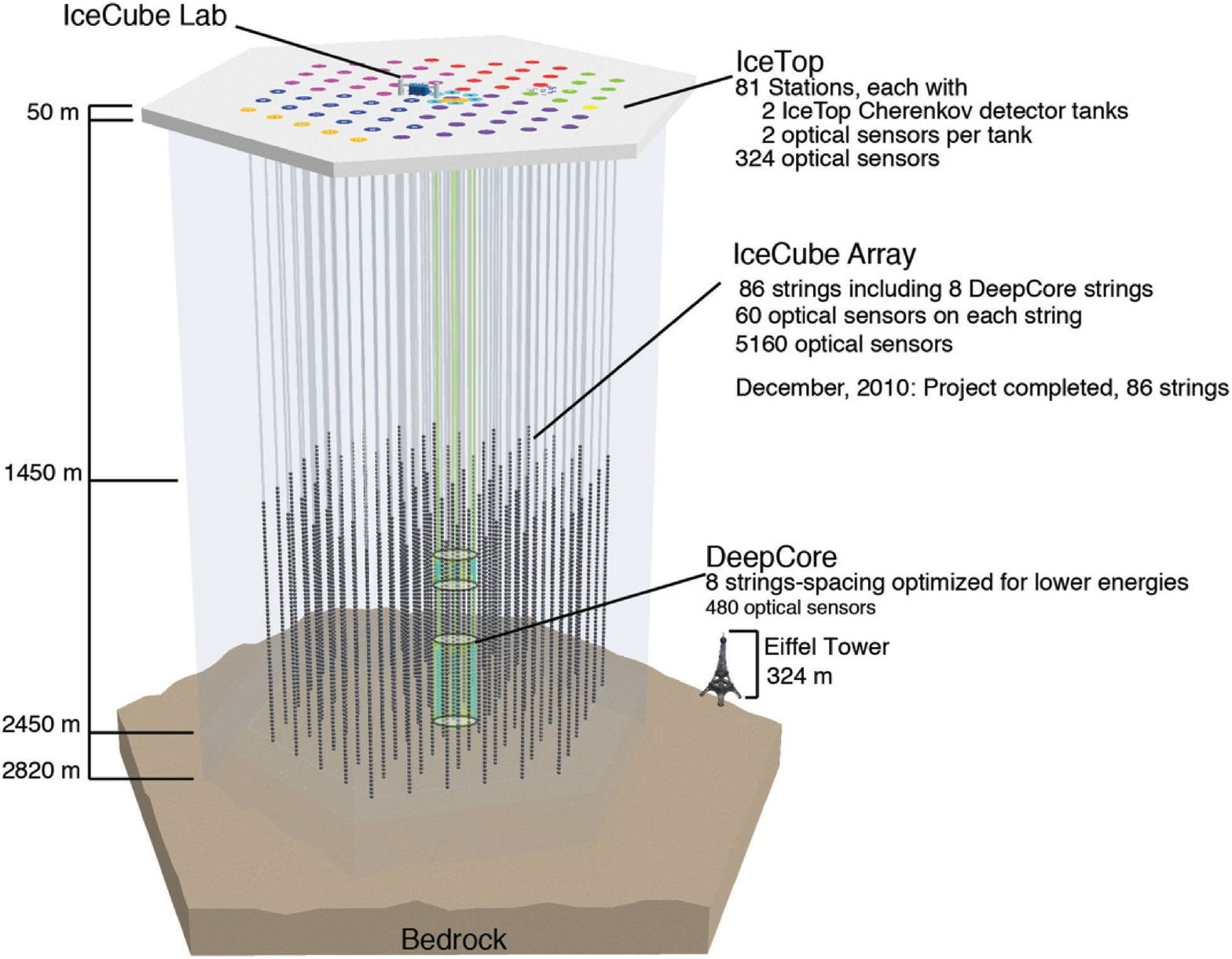}}}
\caption{A schematic view of the IceCube Observatory with the surface array IceTop and the densely instrumented DeepCore.}
\label{fig:icecube}
\end{myfigure}
In the search for high energy neutrinos, the large exposure of IceCube makes it possible to collect an unprecedented number of events in the form of bundles of high energy muons generated in the cosmic ray induced extensive air showers. Although these events represent an overwhelming background in the neutrino searches, they make it possible, for the first time, to determine the degree of anisotropy of cosmic rays from a few TeV to several PeV of particle energy. The persistence of a cosmic ray anisotropy at high energy raises the question of the responsible mechanism. The notion that cosmic ray anisotropy might be connected to the distribution of nearby and recent supernovae is intriguing, and might thus provide a new probe into the origin of the cosmic rays. On the other hand the complex energy-dependent topology suggests that non-diffusive processes in the local interstellar medium most probably play an important role.


\section{Physics Results}
\label{sec:phys}

If the signals from detected Cherenkov photons satisfy specific trigger conditions, an event is defined and recorded by the surface data acquisition system. On-line data filtering at the South Pole reduces the event volume to about 10\% of the trigger rate, based on a series of reconstruction and filter algorithms aimed to select events based on directionality, topology and energy~\cite{reco}. The filter makes it possible to transfer data via satellite from the experimental site for prompt physics analyses.

\subsection{Atmospheric neutrinos}
\label{ssec:atm}

Of the events that trigger IceCube, the vast majority are muon bundles produced by the impact of primary cosmic rays in the atmosphere. Only a small fraction of the detected events ($\sim$10$^{-5}$) are muons produced by the charged current interaction of atmospheric muon neutrinos. The easiest way to reject the down-going muon bundle background is to exclusively select well reconstructed up-going events, since these can only be produced by neutrinos crossing the Earth and interacting in the matter surrounding the detector. Depending on the detector configuration and on the specific reconstruction algorithms and event selection utilized, the atmospheric neutrino sample is characterized by a directional resolution of better than 1$^{\circ}$ above 1 TeV. The corresponding resolution in the estimation of the muon energy is about  0.2-0.3 (in log10 of the energy) for crossing track-like events, and about 0.1 or better for contained cascade-like events. Typically, 30\%-40\% of the up-going events survive the selection with a background contamination of less than about 1\% (see Tab.~\ref{tab:rates}).
{\footnotesize
\begin{mytable}
\caption{Mean rate of muon bundles and atmospheric neutrinos after final event selection for different string configurations of the IceCube Observatory (numbers in italic are predictions).}
\label{tab:rates}
\bigskip
\centerline{\begin{tabular}{|l|l|l|l|}
\hline
strings & year & mean $\mu$ rate & final $\nu_{\mu}$ rate\\
\hline
\hline
22 & 2007 & 500 Hz & 18 / day\\
\hline
40 & 2008 & 1100 Hz & 40 / day\\
\hline
59 & 2009 & 1700 Hz & 130 / day\\
\hline
79 & 2010 & 2000 Hz & {\it 170 / day}\\
\hline
86 & 2011 & 2100 Hz & {\it 200 / day}\\
\hline
\end{tabular}}
\end{mytable}
}
The atmospheric neutrino sample collected by IceCube over the years is the largest ever recorded and currently reaches energies near 400 TeV (see Fig.~\ref{fig:atmo}). For the first time the precision of this measurement is providing a powerful tool to constrain the effects of high energy hadronic interaction models that represent our present knowledge of the cosmic ray induced extensive air showers and the spectrum and composition of primary cosmic rays~\cite{fedynitch}.
\begin{myfigure}
\centerline{\resizebox{60mm}{!}{\includegraphics{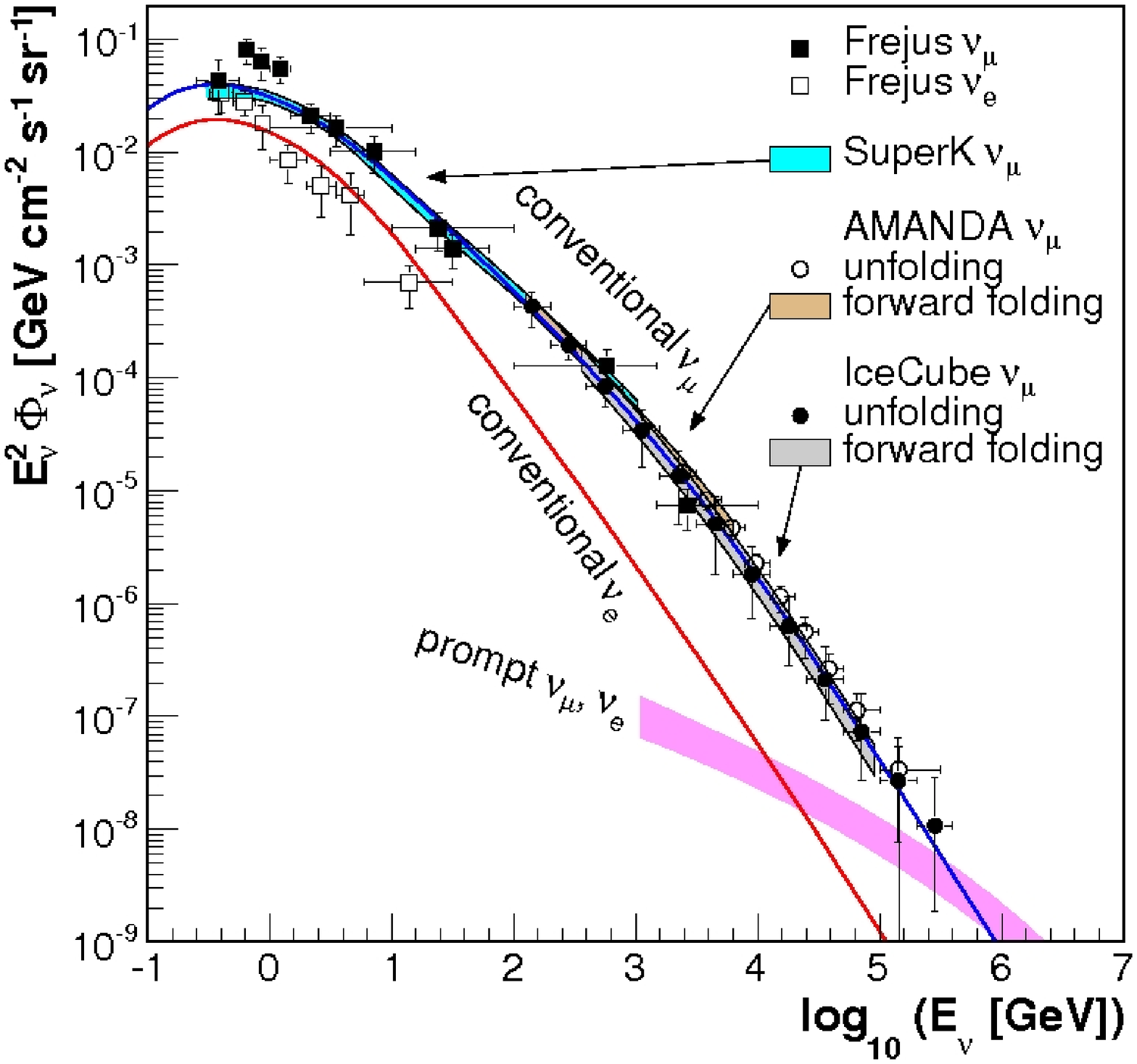}}}
\caption{Collection of theoretical calculations and experimental measurements of the atmospheric neutrino spectrum. Shown is the predicted conventional $\nu_{\mu}+\bar{\nu}_{\mu}$ (blue line) and $\nu_e+\bar{\nu}_e$ (red line) flux from~\cite{honda}, and the predicted prompt flux of neutrinos (magenta band) from~\cite{enberg}. The unfolded energy spectrum~\cite{ic40unfold} (black filled circles) and forward folded spectrum~\cite{ic40ffold} (gray band) from the 40-string IceCube configuration, unfolded spectrum~\cite{amandaunfold} (black open circles) and forward folded spectrum~\cite{amandaffold} (ecru band) from AMANDA are presented. The results from Super-K~\cite{superk} (aqua band) and that from Fr\'ejus~\cite{frejus} (black filled squares for $\nu_{\mu}+\bar{\nu}_{\mu}$ and black open squares for $\nu_e+\bar{\nu}_e$) are also presented.}
\label{fig:atmo}
\end{myfigure}

\subsection{Search for astrophysical $\nu$'s}
\label{ssec:astro}

Atmospheric neutrinos represent an irreducible background for the search of high energy astrophysical neutrinos. If hadronic acceleration is the underlying process of high energy cosmic ray production and $\gamma$ ray observations in galactic and extra-galactic sources, the charged mesons could produce enough neutrinos to be observed in a detector the size of IceCube. 
Fig.~\ref{fig:point} shows the sensitivity (90\% CL) of IceCube for the full-sky search of steady point sources of E$^{-2}$ muon neutrinos as a function of declination, along with that of other experiments. The extension of the point source search to the southern hemisphere is made possible by a high energy event selection that rejects the background down-going events by five orders of magnitude, and restricts neutrino energies to above 100 TeV. Still dominated by high energy large muon bundles, this makes the southern hemisphere poor in atmospheric neutrinos yielding a low neutrino detection sensitivity. Nevertheless, this provides IceCube with a full-sky view that complements coverage of the neutrino telescopes in the Mediterranean. The figure shows the sensitivities from IceCube and other observatories (interpreted as the median upper limit we expect to observe from individual sources across the sky) along with upper limits from selected sources. The sensitivity is reaching the level of current predictions for flux from astrophysical sources ({\it i.e.} below 10$^{12}\times$ E$^{-2}$ TeV cm$^{-2}$ s$^{-1}$) although the discovery potential, defined to be 5$\sigma$ for 50\% of the trials, is typically a factor of three higher than the sensitivity. Therefore constraints on the parameters of hadronic acceleration models are starting to develop.
\begin{myfigure}
\centerline{\resizebox{60mm}{!}{\includegraphics{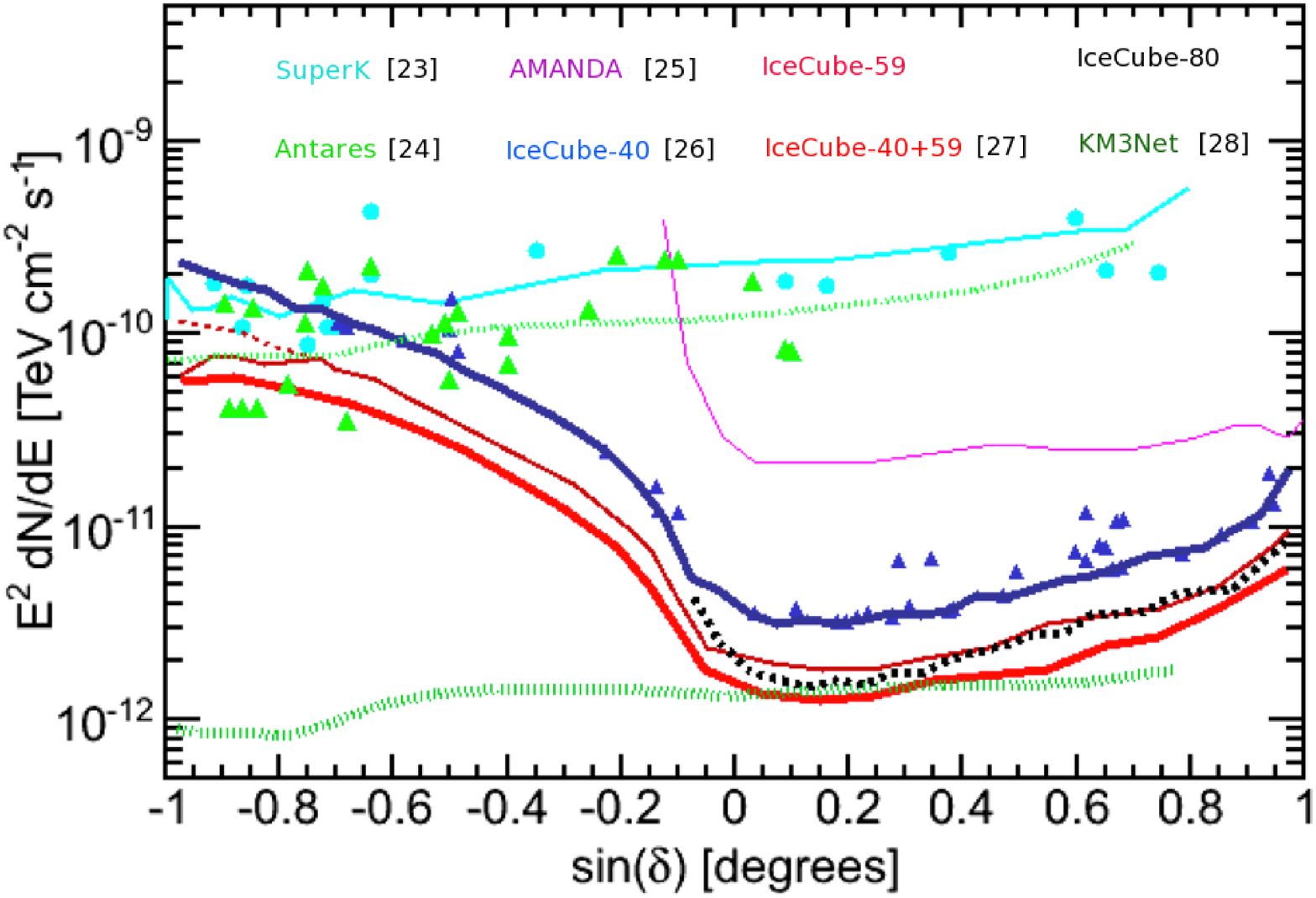}}}
\caption{Sensitivity (90\% CL) for a full-sky search of steady point sources of muon neutrinos with an E$^{-2}$ energy spectrum as a function of declination angle for IceCube and other experiments. Note that for IceCube, events with $\delta<$ 0$^{\circ}$ are down-going, coming from the southern hemisphere, and events with $\delta>$0$^{\circ}$ are up-going and come from the northern hemisphere. 
}
\label{fig:point}
\end{myfigure}
\begin{myfigure}
\centerline{\resizebox{70mm}{!}{\includegraphics{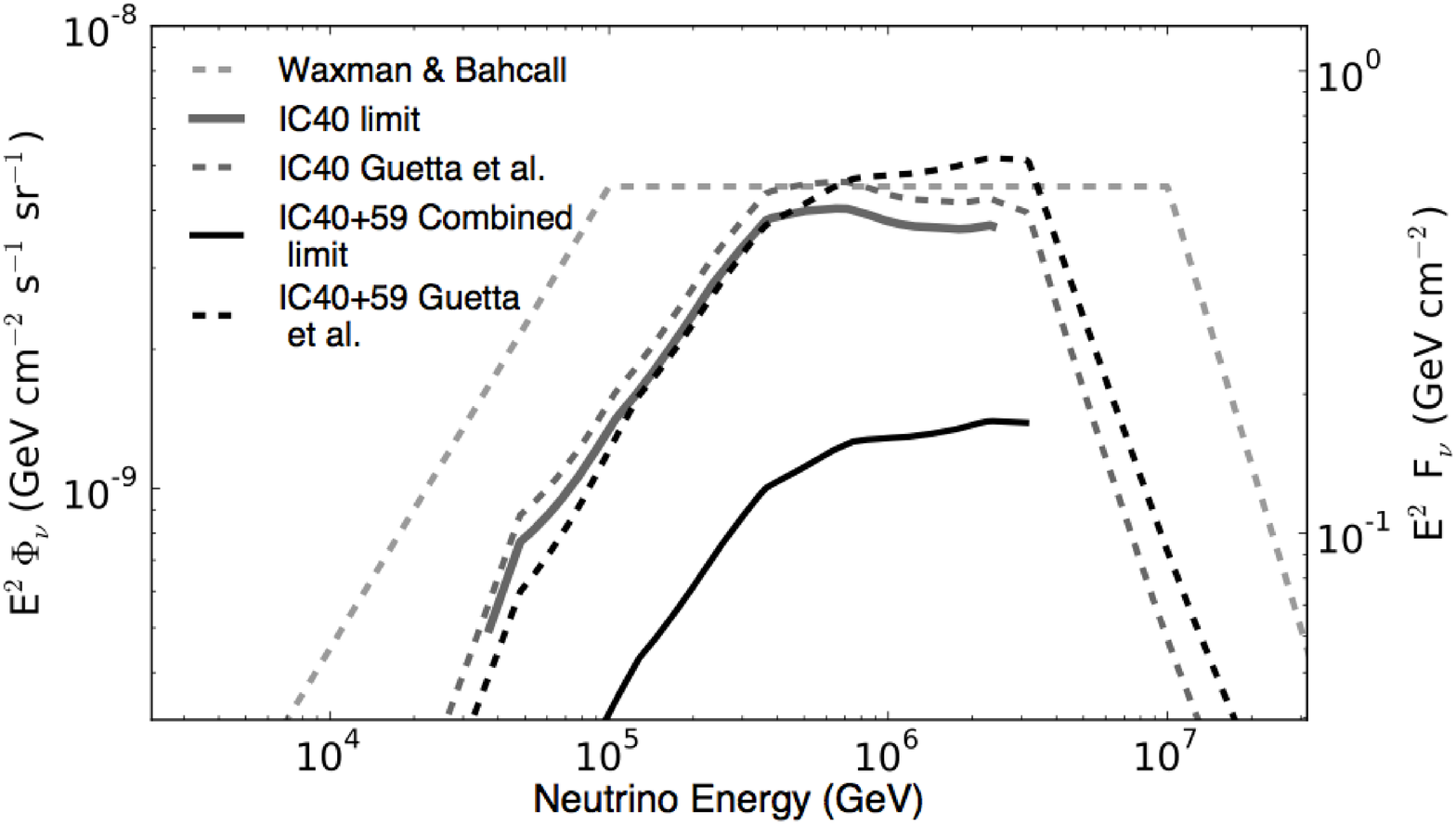}}}
\caption{Upper limits (90\% CL) for neutrino searches in coincidence with Gamma Ray Bursts with 40 strings of IceCube and the combined 40- and 59-string detector configurations~\cite{icgrb}. Also shown is the Waxman \& Bahcall predicted average flux~\cite{wbgrb}.}
\label{fig:grb}
\end{myfigure}
Searches for neutrinos from transient~\cite{ictransient} and periodic~\cite{icperiodic} sources have also been performed. In particular, a time window scan for transient sources (with no external triggers) shows that the discovery potential drops by a factor of 2 if searching for 1 day duration flares. A particular search for transient sources is that for neutrinos from GRB. For the first time, the IceCube Observatory has provided a definitive test of the GRB models with the most stringent constraints. Fig.~\ref{fig:grb} shows the upper limits obtained with the data collected by the 40-string configuration of IceCube and by the combined data of the 40- and 59-string configurations~\cite{icgrb}. For each detector configuration, a list of GRBs detected during the corresponding physics runs was compiled and the predicted neutrino flux was calculated based on the $\gamma$ ray spectrum shown in~\cite{guetta}. The corresponding stacked neutrino flux was used to search for events collected within the time window in which 5\% to 95\% of the fluence is recorded ({\it i.e.} $T_{90}$). The upper limit is about 3 times below the predicted flux of the Waxman \& Bahcall model, challenging the hypothesis that GRB are the sources of Ultra High Energy Cosmic Rays (UHECR). This result has profound consequences for the predicted flux of neutrinos produced by the interaction of UHECR with the cosmic microwave background, the so-called cosmogenic neutrinos, as well as for the GeV-TeV $\gamma$ ray background flux (see for instance~\cite{gzk,ahlers}). It is important to note that it was recently shown that the fireball model with refined assumptions yields a 10 times smaller predicted flux (see~\cite{baerwald,huemmer}).
%
%
There is the possibility that the bulk of cosmic rays does not originate from individual sources, but from large-scale acceleration processes in superbubbles or even Galaxy clusters. In addition, unresolved sources of cosmic rays over cosmological times are expected to have produced detectable fluxes of diffuse neutrinos. Since shock acceleration is expected to provide an $\sim$E$^{-2}$ energy spectrum, harder than the $\sim$E$^{-3.7}$ of the atmospheric neutrinos, the diffuse flux is expected to dominate at high energy where the sensitivity is strongly dependent on the experimental quality of the selected events.
Fig. \ref{fig:diff} shows a collection of sensitivities and upper limits (90\% CL) for an E$^{-2}$ flux of $\nu_{\mu}+\bar{\nu}_{\mu}$, from AMANDA, Antares and various IceCube configurations compared to the experimental and theoretical flux of the atmospheric neutrinos and various models of astrophysical neutrinos. The most recent results lie below the Waxmann \& Bahcall neutrino bound~\cite{wb}, again indicating IceCube's potential for discovering the origin of cosmic rays.

In the Ultra High Energy range (UHE), above $\sim$10$^6$ GeV, IceCube is reaching a competitive sensitivity as well. At this level one begins to reach current models of cosmogenic neutrino production (see Fig.~\ref{fig:ehe}) that are simultaneously constrained by the current observations of UHECRs and the GeV $\gamma$ rays by Fermi-LAT~\cite{ahlers2}. Taking into account that UHECR mass composition is a key ingredient for the absolute flux and spectral shape of cosmogenic neutrinos~\cite{gzk}, its large uncertainty still weighs profoundly on current models. This means that although the IceCube sensitivity to UHE neutrinos is currently the best ever achieved below 10$^{10}$ GeV it might be still far from the actual flux. From this point of view, the current developments toward a radio array in Antarctica, such as Askaryan Radio Array (ARA)~\cite{ara} is a natural extension toward the highest energies.
\begin{myfigure}
\centerline{\resizebox{50mm}{!}{\includegraphics{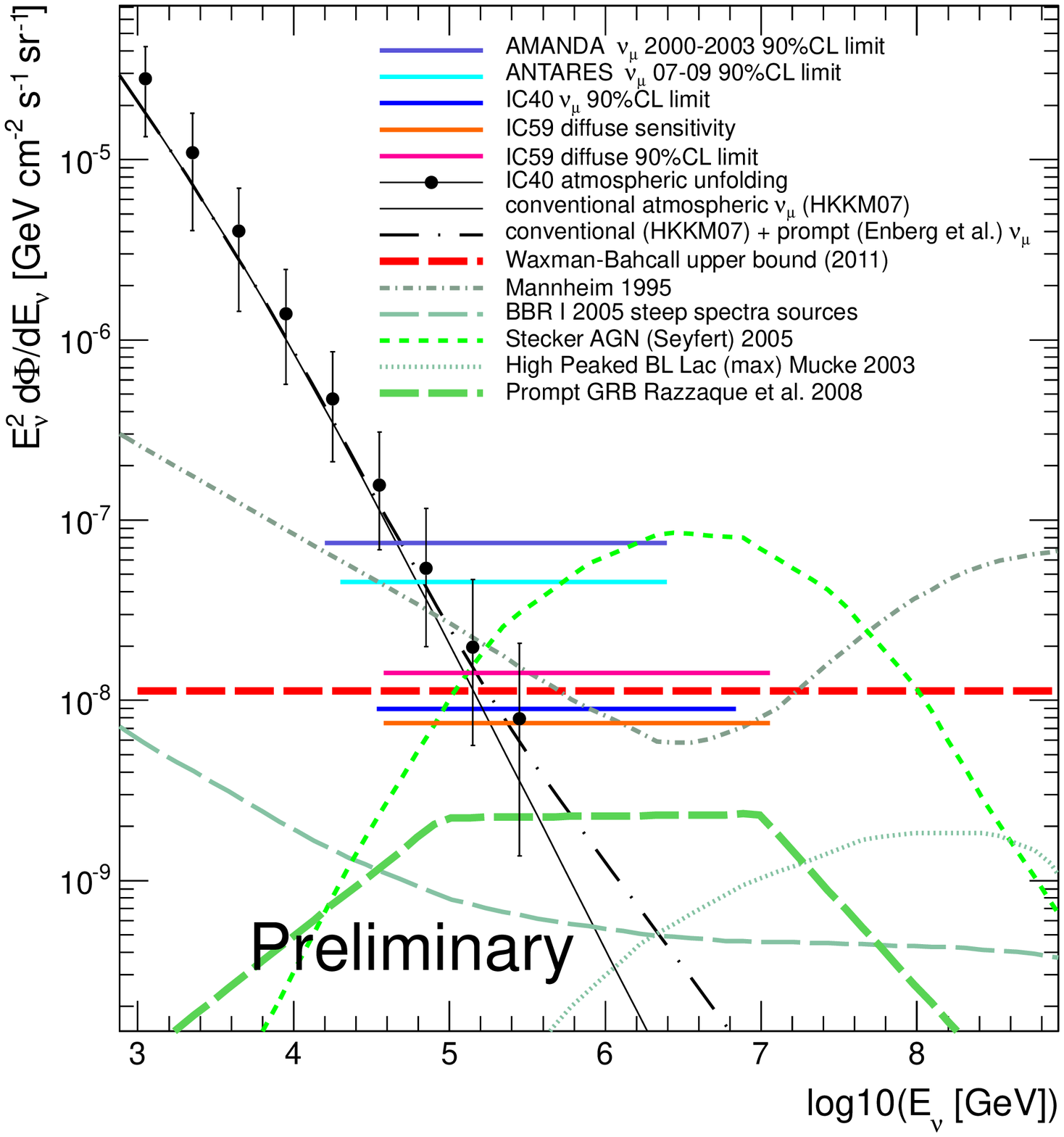}}}
\caption{Experimental upper limits (90\% CL) for the diffuse muon neutrino flux (including the preliminary result from the 59-string configuration of IceCube) along with atmospheric neutrino observations and theoretical models of atmospheric and extra-terrestrial neutrino fluxes. From top to bottom in the legend~\cite{amandadiffuse, antaresdiffuse, ic40diffuse, ic40unfold, honda, wb, wb2, agn, bllacs, becker, stecker}
}
\label{fig:diff}
\end{myfigure}
\begin{myfigure}
\centerline{\resizebox{50mm}{!}{\includegraphics{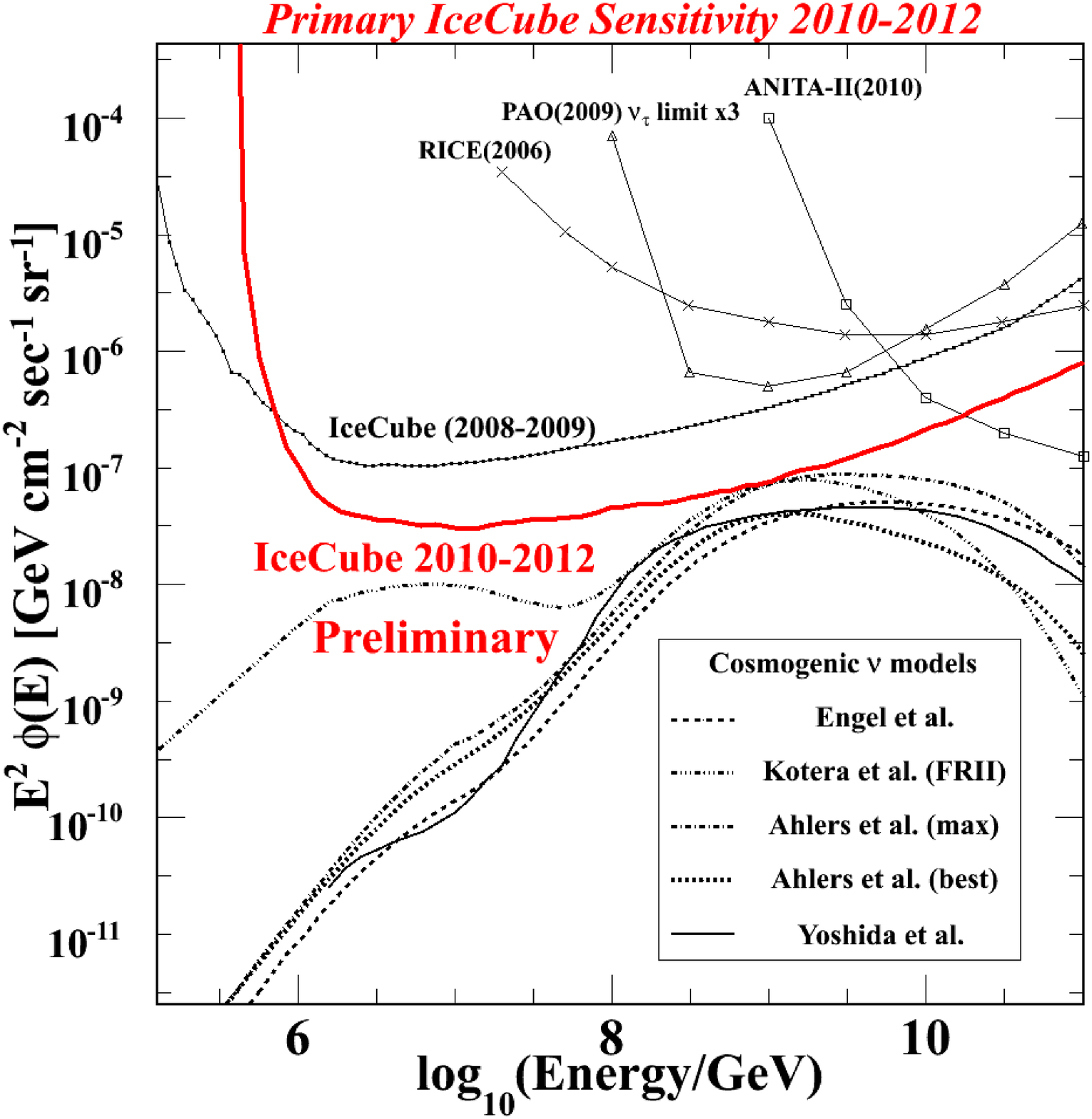}}}
\caption{Preliminary sensitivity (90\% CL) for the detection of UHE neutrinos, compared to other experimental results and to predictions~\cite{kotera, ahlers2, yoshida}. The sensitivity curves are evaluated at each decade of energy.}
\label{fig:ehe}
\end{myfigure}
It is worth noting that the preliminary sensitivity for an arbitrary spectrum, shown in Fig.~\ref{fig:ehe}, has a minimum just above 1 PeV, where no significant cosmogenic neutrino flux is expected. In the experimental analysis performed on data collected during 2010-12, where events with a large number of detected photons were selected, two events were found on a background of conventional atmospheric neutrinos of 0.3. The events deposited an energy in the detector of about 1 PeV, and further study is underway to determine their nature. One possible hypothesis is that these events represent an upper fluctuation of the prompt neutrino production in the atmosphere from the decay of heavy charm mesons.

\subsection{Cosmic ray anisotropy}
\label{ssec:anyso}

The large number of muon bundle events collected by IceCube (about 10$^{10}$-10$^{11}$ each year, depending on the detector configuration) makes it possible to study the arrival direction distribution of the cosmic rays at a level of about 10$^{-5}$. The bundles of highly collimated atmospheric muons share the same direction as the parent cosmic ray particle. Since this study does not require highly well reconstructed muon directions, all collected and reconstructed events with a median angular resolution of about 3$^{\circ}$ are used. Using full simulation of cosmic ray induced extensive air shower we find that the median particle energy of the IceCube data sample is about 20 TeV. With these data IceCube provides the first high statistics determination of the anisotropy of galactic cosmic rays in the southern hemisphere in the multi-TeV energy range.


The large scale anisotropy observed by IceCube~\cite{anisotropylarge10} appears to complement the observations in the northern hemisphere, providing for the first time an all-sky view of TeV cosmic ray arrival directions. The sky map obtained by subtracting an averaged map (over a scale of 30$^{\circ}$-60$^{\circ}$) from the data~\cite{anisotropysmall11}, shows significant small angular scale structures in the cosmic ray anisotropy, similarly to observations in the northern hemisphere~\cite{abdo2, argo}.
Another interesting result obtained by IceCube is the persistence of the anisotropy at an energy in excess of 100 TeV. At such energies a different structure is observed that can be interpreted in terms of a different phase~\cite{anisotropylarge12} as already reported by the EAS-TOP shower array in the northern hemisphere for the first time~\cite{eastop}. The observation at high energy was recently confirmed by the preliminary result from the IceTop shower array~\cite{icetopmarcos}. The change of the anisotropy pattern at about 100 TeV may suggest that the heliosphere could have an effect in flipping the apparent direction of the anisotropy. In fact, at about 100 TeV the cosmic rays' gyro-radius in the 3 $\mu$G local interstellar magnetic field is of the order of magnitude of the elongated heliosphere. Below this energy scale the scattering processes on the heliospheric perturbations at the boundary with the interstellar magnetic field might be the dominant processes affecting the global cosmic ray arrival distribution and the small angular structure as well (see~\cite{scattering} where a review of other proposed models is also given). The Milagro observation of a likely harder than average cosmic ray spectrum from the localized excess region toward the direction of the heliotail, the so-called region B in~\cite{abdo2} and also observed by ARGO-YBJ shower array~\cite{argo}, have triggered astrophysics interpretations (see ~\cite{astro1, astro2, astro3}). However, this may suggest that some type of re-acceleration mechanism associated with cosmic ray propagation in the turbulent heliospheric tail might occur~\cite{reconnection1, reconnection2}. On the other hand, the TeV cosmic ray anisotropy is a tracer of the local interstellar magnetic field, and it might indicate cosmic ray streaming along the magnetic field lines due to the Loop I shell expanding from the Scorpion-Centaurus Association~\cite{frisch12}.

If the local propagation effects on the cosmic ray anisotropy below 100 TeV are dominant, at higher energy it is reasonable to believe that the persistent anisotropy might be a natural consequence of the stochastic nature of cosmic ray galactic sources, in particular nearby and recent SNRs~\cite{erlykin, blasi, biermann}.



\end{multicols}
\end{document}